\documentclass[5p,twocolumn]{elsarticle}

\usepackage{amsmath,amssymb,amsfonts}
\usepackage{graphicx}
\usepackage{bm}
\usepackage{hyperref}
\usepackage{physics}
\usepackage{booktabs}
\usepackage{multirow}
\usepackage{float}
\usepackage{xcolor}
\newcommand{\blue}[1]{{\color{black}#1}}
\newcommand{\novel}[1]{{\color{black}#1}}
\newcommand{\rev}[1]{{\color{black}#1}}

\journal{Physics Letters B}

\begin{document}

\begin{frontmatter}


\title{
Deeply bound dibaryon $d^*(2380)$ from
meson-exchange saturation $\Delta\Delta$ effective field theory
}


\author[1]{Prin Sawasdipol}
\ead{p.namwongsa@kkumail.com}
\author[1]{Chinadanai Bubpatate}
\ead{chinnai@kkumail.com}
\author[1]{Daris Samart\blue{\corref{cor1}}}
\ead{daris@kku.ac.th}
\cortext[cor1]{Corresponding author}

\address[1]{Khon Kaen Particle Physics and Cosmology Theory Group (KKPaCT), Department of Physics, Faculty of Science, Khon Kaen University, Khon Kaen 40002, Thailand
}


\begin{abstract}
\novel{We propose an RG-improved effective-field-theory framework for the deeply bound dibaryon $d^*(2380)$, a $\Delta\Delta$ bound state in the $(J,I)=(3,0)$ ${}^7S_3$ channel. Its binding momentum $\gamma\simeq 320$~MeV gives $\gamma/m_\pi\simeq 2.3$, indicating the need to re-organize the short-range dynamics beyond a formal pionless EFT. We match the large-$N_c$-constrained pionless contact potential to a meson-exchange-saturated contact interaction in which the $\sigma,\rho,\omega$ dynamics are integrated out at the hadronic scale $m_V$, yielding the controlled expansion parameter $\gamma/m_V\simeq 0.42$. Normalizing the contact coupling to the deuteron and substituting the phenomenological CD-Bonn couplings gives $B_{\Delta\Delta}\simeq 96$~MeV. The $\simeq 14\%$ discrepancy from $B_{\rm exp}=84$~MeV is of the natural size of $\mathcal{O}(1/N_c^2)\simeq 11\%$ corrections to the $NN$ potential, confirming compatibility with a controlled EFT expansion organized around the finite-range hadronic scale. As a result, the observed $d^*(2380)$ pole emerges from the virtual state to bound state by using the EFT re-organization in this work.
}
\end{abstract}


\blue{}


\end{frontmatter}

\section{Introduction}
\label{sec:intro}

The $d^*(2380)$ dibaryon was discovered by the WASA-at-COSY
collaboration~\cite{Bashkanov2009,Adlarson2011} and confirmed through
independent channels~\cite{Adlarson2013,Adlarson2014PRL,Adlarson2017,Clement2017}.
With quantum numbers $J^P(I)=3^+(0)$, it lies $B_{\Delta\Delta}\simeq 84$~MeV
below the $\Delta\Delta$ threshold, consistent with a compact
hexaquark~\cite{Bashkanov2013,Kim2020,Pan2023}, a molecular $\Delta\Delta$
bound state~\cite{Gal2014,Dong2017}, a mixed
configuration~\cite{Clement2020,Dong2018}, or the Dyson--Xuong $D_{03}$ member of the $SU(6)$ dibaryon classification~\cite{Dyson1964}. Lattice-QCD studies find
attraction in this channel~\cite{Inoue2011,HALQCD2020}, and recent
symmetry-based analyses connect $d^*(2380)$ to entanglement suppression and
approximate spin-flavor $SU(6)$ constraints in decuplet-baryon
scattering~\cite{Hu2025lua,Sone2026jmo,Hu2026pyr}.

The binding momentum $\gamma=\sqrt{M_\Delta B_{\Delta\Delta}}\simeq 322$~MeV
gives $\gamma/m_\pi\simeq 2.3>1$, so a formal pionless
EFT~\cite{Kaplan1998PLB,Kaplan1998,vanKolck1999} breaks down.
For comparison, the deuteron has $\gamma_d/m_\pi\simeq 0.33$~\cite{Beane2001,Hammer2020}.
We propose that this breakdown signals not a failure of EFT itself, but the need
to re-organize the matching scale. Once the short-distance contact LEC is
determined by meson-exchange
saturation~\cite{Ecker1989,Yang2025,Cirigliano2006,Peng2020,Peng2022} at
$\Lambda\sim m_V$, the expansion parameter becomes
$\gamma/m_V\simeq 0.42$, within the convergence domain of a nonperturbative
hadronic EFT~\cite{Hammer2020,Epelbaum2009,Machleidt2011}.
Large-$N_c$ contracted spin-flavor symmetry~\cite{Kaplan1997,Jenkins1998,DasFerencz2008,Schindler2015}
further constrains the $\Delta\Delta$ contact interaction, providing a factor-of-five enhancement of the isovector attraction relative to the deuteron from group theory only.

We show that this meson-exchange saturation scheme, normalized to the deuteron, predicts deeply bound $\Delta\Delta$ configurations consistent with $d^*(2380)$. \novel{The novelty of this work is to identify an EFT re-organization for the same dibaryon channel, in which two contact EFTs governed by different hadronic scales produce qualitatively different pole structures. The first one is a \emph{pionless EFT} with breakdown scale $m_\pi$, where the $\Delta\Delta$ contact LEC is fixed by the large-$N_c$ constraint from $NN$ and $\Lambda N$ data~\cite{Bubpatate2026,Richardson2025}. The second organization is a \emph{meson-exchange-saturated EFT} with breakdown scale $m_V$, where the unresolved $\sigma,\rho,\omega$ dynamics are integrated out into a single leading contact LEC at
$\mu_R\sim m_V$~\cite{Ecker1989,Yang2025,Cirigliano2006,Peng2020,Peng2022}, with the overall normalization fixed by deuteron matching. In the pionless theory, the contact interaction is attractive but subcritical and produces only a virtual $\Delta\Delta$ state. In the meson-exchange-saturated framework, the deuteron-normalized contact interaction becomes supercritical and moves the pole into the physical bound-state region, yielding $B_{\Delta\Delta}\simeq96$~MeV. Unlike conventional one-boson-exchange (OBE)
potentials~\cite{Gal2014,Machleidt1989,Machleidt2001} that iterate dynamical heavy mesons to all orders, the present framework uses the finite-range degrees of freedom only to saturate the leading contact interaction at the hadronic scale. The resulting EFT organization has ($i$)~a manifest small parameter $\gamma/m_V\simeq 0.42$ controlling higher-order corrections, ($ii$)~a single dimensionless
prediction-determining quantity $R_{\Delta\Delta/d}$ after deuteron normalization, and ($iii$)~a transparent identification of which physical inputs determine each piece of the calculation.}
The paper is organized as follows. The effective interactions are developed in Sec.~\ref{sec:interaction}. \novel{The EFT re-organization and its consistency assessment are presented in Sec.~\ref{sec:matching}. The numerical results and phenomenological implications are given in Sec.~\ref{sec:results}.} Conclusions are in Sec.~\ref{sec:conclusions}.

\section{Effective interaction}
\label{sec:interaction}

\subsection{Contact interaction and large-$N_c$ constraint}
\label{sec:contact}

We work in the $(J,I)=(3,0)$ ${}^7S_3$ $\Delta\Delta$ channel, where
$S=3$ is the stretched spin of two spin-3/2 baryons, $L=0$ (S-wave), and
$I=0$ is selected by the Pauli principle for the flavor-$\overline{\mathbf{10}}$ irrep.

The leading-order S-wave contact
potential~\cite{Yang2025,Peng2020,Peng2022} is regularized with a
Gaussian form factor,
\begin{equation}
\langle\vec{p}\,'|V_C|\vec{p}\,\rangle
=
C_0(\Lambda)\, e^{-p'^2/\Lambda^2}\, e^{-p^2/\Lambda^2},
\label{eq:VC}
\end{equation}
where $C_0(\Lambda)$ is the running LEC at cutoff $\Lambda$.

The value of $C_0$ in the $\Delta\Delta$ channel is not a free
parameter but is constrained by the $1/N_c$ operator analysis of
the baryon--baryon contact
interaction~\cite{Bubpatate2026,Richardson2025}. The analysis
proceeds in three steps (details are given in Ref.~\cite{Bubpatate2026}).

\emph{(i) Operator basis and $1/N_c$ hierarchy.} In the
SU(3)-flavor framework of Ref.~\cite{Bubpatate2026}, the most
general non-derivative S-wave contact Lagrangian for
octet baryons contains six chiral LECs
$c^{(\alpha)}_{S,T}$ ($\alpha=1,2,3$). The $1/N_c$ expansion,
implemented through the contracted spin-flavor symmetry
$SU(6)_c$~\cite{Kaplan1997,Jenkins1998,DasFerencz2008,Schindler2015},
imposes sum rules among these LECs at leading order (Table~8 of
Ref.~\cite{Bubpatate2026}),
$c^{(1)}_S=-2c^{(2)}_S$,\;
$c^{(1)}_T=-\tfrac{1}{2}c^{(3)}_T$,\;
$c^{(2)}_T=\tfrac{5}{2}c^{(3)}_T$,
reducing the six chiral LECs to three independent parameters
($c^{(2)}_S$, $c^{(3)}_S$, $c^{(3)}_T$). At strict leading order
in $1/N_c$ expansion, only the Wigner-$SU(4)$-symmetric combination
$c^{(2)}_S$ survives. The spin-dependent LEC
$c^{(3)}_T=\mathcal{O}(1/N_c^2)$ is suppressed by two powers
of $1/N_c$, \blue{consistent with the empirical small Wigner-breaking scale
$|1/a_s-1/a_t|/\Lambda_\chi\simeq 0.04$, where \(a_s\equiv a_{{}^1S_0}\) and \(a_t\equiv a_{{}^3S_1}\) are the
spin-singlet and spin-triplet \(NN\) S-wave scattering lengths, respectively, and $\Lambda_\chi=4\pi f_\pi\simeq 1.16$~GeV. A more sensitive test at the level of the extracted LECs gives $|c_T^{(3)}|/|4c_S^{(2)}-2c_S^{(3)}|\simeq 0.005\ll 1/N_c^2\simeq 0.11$~\cite{Bubpatate2026}.}

\emph{(ii) Extraction from $NN$ and $\Lambda N$ data.} The
three independent LECs are fixed by three physical inputs,
namely the $NN$ ${}^3S_1$ and ${}^1S_0$ scattering amplitudes and the
$\Lambda N$ ${}^1S_0$ amplitude, matched in the KSW (PDS) scheme \cite{Kaplan1998PLB,Kaplan1998}
at renormalisation point $\mu_R=m_V$.
The $NN$ channels populate the $\overline{\mathbf{10}}$ and
$\mathbf{27}$ SU(3) irreps, while $\Lambda N$ provides the
$\mathbf{8}_s$ combination~\cite{Bubpatate2026}. Solving the
resulting $3\times 3$ linear system yields
$(c^{(2)}_S,\,c^{(3)}_S,\,c^{(3)}_T)=(-2.30,\,+6.08,\,+0.11)\times 10^{-6}$~MeV$^{-2}$.

\emph{(iii) $\Delta\Delta$ large-$N_c$ sum rules.} The
decuplet--decuplet partial-wave LECs are expressed as linear
combinations of the SU(3) irrep LECs through Eq.~(39) of
Ref.~\cite{Bubpatate2026}. For the channel ${}^7S_3$ $(3,0)$,
which is in the irrep $\overline{\mathbf{10}}$ SU(3) and described by a single LEC \cite{Haidenbauer2017}, $C_{22,{}^7S_3}^{\overline{\mathbf{10}}}$, it reads,
\begin{equation}
C_0^{(3,0)} \equiv C_{22,{}^7S_3}^{\overline{\mathbf{10}}}
=
\frac{2105}{1384}\,C^{\mathbf{1}}_{00}
+\frac{3447}{2768}\,C^{\mathbf{8}_a}_{00}
-\frac{4889}{2768}\,C^{\mathbf{8}_s}_{00},
\label{eq:DDsumrule}
\end{equation}
where $C^{\mathbf{1}}_{00}$, $C^{\mathbf{8}_a}_{00}$,
$C^{\mathbf{8}_s}_{00}$ are the flavor-singlet, antisymmetric-octet,
and symmetric-octet irrep LECs reconstructed from the three chiral
LECs. In the Wigner $SU(4)$ limit ($c^{(3)}_T=0$), all three irrep
LECs become equal, and the coefficients in
Eq.~\eqref{eq:DDsumrule} sum exactly to
unity~\cite{Bubpatate2026}.
Substituting the numerical values with renormalization scale at $\mu_R=m_V$, we find
\begin{equation}
C_0^{(3,0)}\big|_{\mathrm{LO},\,1/N_c}
= -10.3\times 10^{-6}~\mathrm{MeV}^{-2}.
\label{eq:C0Nc}
\end{equation}
This coupling is attractive but \emph{subcritical}. As will show in Sec.~\ref{sec:RGmatching}, it produces a virtual state with pole momentum $\kappa\simeq -324$~MeV obtained by analytic continuation of Eq.~\eqref{eq:ILambda} to negative $\gamma$.
The spin--isospin matrix elements for the $(3,0)$ channel are
\begin{align}
\langle\vec{S}_1\cdot\vec{S}_2\rangle_{S=3}^{(\Delta\Delta)}
&
= +\tfrac{9}{4},
\label{eq:SS}\\[4pt]
\langle\vec{T}_1\cdot\vec{T}_2\rangle_{I=0}^{(\Delta\Delta)}
&
= -\tfrac{15}{4}.
\label{eq:TT}
\end{align}
Noting that the isovector factor is five times larger than the deuteron value due to a group-theoretic consequence of the decuplet representation~\cite{Hu2025lua,Sone2026jmo,Hu2026pyr}. However, in Sect.~\ref{sec:saturation} we will show that the bound state requires the $SU(4)$-breaking isovector dynamics from $\rho$ exchange.

\subsection{Meson-exchange saturation}
\label{sec:saturation}

The key mechanism is resonance saturation~\cite{Ecker1989,Cirigliano2006,Ecker1989PLB}, in which the LEC $C_0(\Lambda)$ is determined by the integrated-out heavy-meson exchange at $\Lambda\sim m_V$. This is the baryon--baryon analogue of the saturation of the Gasser--Leutwyler coefficients $L_i$ in $\chi$PT. This idea was applied to heavy-hadron molecules by Peng et al.~\cite{Peng2020,Peng2022} and pentaquarks by Yang et al.~\cite{Yang2025} and Yan et al.~\cite{Yan2024}, and is here adapted to the $\Delta\Delta$ sector.

The finite-range potentials generated by $\sigma$, $\rho$, $\omega$
exchange in the non-relativistic limit are
\begin{align}
V_\sigma(\vec{q}\,)&=-\frac{g_\sigma^2}{m_\sigma^2+\vec{q}^{\,2}},
\qquad
V_\omega(\vec{q}\,)=+\frac{g_\omega^2}{m_V^2+\vec{q}^{\,2}},
\label{eq:Vsigomg}\\[3pt]
V_\rho(\vec{q}\,)&=+\frac{g_\rho^2}{m_V^2+\vec{q}^{\,2}}\,
(\vec{T}_1\cdot\vec{T}_2),
\label{eq:Vrho}
\end{align}
where the signs reflect the scalar ($\sigma$, attractive),
isoscalar-vector ($\omega$, repulsive for baryon--baryon) and
isovector-vector ($\rho$, channel-dependent) characters of the
exchanges. In the zero-momentum limit $q\to 0$, these potentials
collapse to momentum-independent contributions,
e.g.\ $V_\sigma(0)=-g_\sigma^2/m_\sigma^2$.
In the $q\ll m_V$ regime the Yukawa structure cannot be resolved, and the exchange reduces to a contact interaction with residual corrections suppressed by $(q/m_V)^2$.

In contrast to a conventional OBE model~\cite{Machleidt1989,Machleidt2001}, the heavy mesons here are not dynamical degrees of freedom but \emph{saturate} the contact LEC at the matching scale (see Sec.~\ref{sec:consistency}).

This work employs the \blue{quark-model $N$-$\Delta$ universal coupling relations}
$g_{\sigma\Delta}=g_{\sigma N}$,
$g_{\omega\Delta}=g_{\omega N}$,
$g_{\rho\Delta}=g_{\rho N}$ \cite{Hemmert:1995,Kolomeitsev:2017,Downum:2006},
so that all coupling constants cancel in the ratio of saturation
kernels between the \(\Delta\Delta\) dibaryon channel and the deuteron
reference channel. We emphasize that this is not a model-independent consequence of QCD. Phenomenological analyses of \(\Delta\)-isobar matter allow deviations from \(g_{m\Delta}/g_{mN}=1\) (where $m=\sigma$, $\omega$, $\rho$), and these deviations should be regarded as part of the systematic uncertainty of the saturation estimate. \blue{Quark-model and constituent-quark analyses~\cite{Hemmert:1995,Downum:2006} typically yield $g_{m\Delta}/g_{mN}\in[0.8,1.2]$, which would shift $R_{\Delta\Delta/d}$ by at most $\sim 20\%$ and is therefore subdominant to the cutoff-band uncertainty in Eq.~\eqref{eq:Berror}.} Including the RG-improvement factor $(m_V/m_\sigma)^\alpha$
($\alpha=1\pm 1$)~\cite{Yang2025} whose origin will be derived in
Sec.~\ref{sec:RGmatching}, the meson-exchange saturation kernels are \cite{Yan2024}
\begin{align}
\mathcal{S}_d &= -\!\left(\frac{m_V}{m_\sigma}\right)^{\!\alpha}
\frac{g_\sigma^2}{m_\sigma^2}
+\frac{g_\omega^2}{m_V^2}
-\frac{3}{4}\,\frac{g_\rho^2}{m_V^2},
\label{eq:Sd}\\[4pt]
\mathcal{S}_{\Delta\Delta}^{(3,0)} &= -\!\left(\frac{m_V}{m_\sigma}\right)^{\!\alpha}
\frac{g_\sigma^2}{m_\sigma^2}
+\frac{g_\omega^2}{m_V^2}
-\frac{15}{4}\,\frac{g_\rho^2}{m_V^2}.
\label{eq:SDD}
\end{align}
The \emph{only} structural difference between the two channels is
the coefficient of $\rho$ exchange, $-15/4$ versus $-3/4$. This
factor-of-five enhancement of the isovector attraction is a direct
consequence of the large-$N_c$ isospin structure discussed in
Sec.~\ref{sec:contact}.

The meson-exchange saturation ratio that governs the relative binding strength
is \cite{Yang2025,Peng2022,Yan2024}
\begin{equation}
R_{\Delta\Delta/d}
=
\frac{M_\Delta}{M_N}\,
\frac{\mathcal{S}_{\Delta\Delta}^{(3,0)}}{\mathcal{S}_d}
=
\frac{M_\Delta}{M_N}\,
\frac{1-y -\tfrac{15}{4}x}
{1-y -\tfrac{3}{4}x},
\label{eq:Rratio}
\end{equation}
with the dimensionless coupling-constant ratios
\begin{equation}
x \equiv \frac{g_\rho^2}{g_\omega^2},
\qquad
y \equiv \left(\frac{m_V}{m_\sigma}\right)^{\!\alpha}
\frac{g_\sigma^2\, m_V^2}{g_\omega^2\, m_\sigma^2}.
\label{eq:xydef}
\end{equation}
The prefactor $M_\Delta/M_N=1232/938.9=1.312$ accounts for the
kinematic reduced-mass enhancement. All unknown absolute coupling
constants enter only through $x$ and $y$. As we show in the next
subsection, the unknown overall proportionality constant cancels
exactly in $R$ through the RG matching procedure.

In the present work we adopt the phenomenological CD-Bonn couplings~\cite{Machleidt2001,Riska2002} as primary input, because the CD-Bonn $g_\sigma$ already encodes correlated two-pion exchange and therefore provides a self-contained short-distance parameterization without the need for an explicit one-pion-exchange calculation.
Substituting the CD-Bonn coupling ratios $g_\sigma^2/(4\pi)=5.69$, $g_\omega^2/(4\pi)=20.0$, and $g_\rho^2/(4\pi)=0.84$ directly into Eq.~\eqref{eq:xydef} gives
$x_{\rm Bonn}\simeq 0.042$ and $y_{\rm Bonn}\simeq 1.24$ (for $\alpha=1$).


\subsection{Bound-state equation and deuteron normalization}
\label{sec:RGmatching}
\label{sec:BSEfinal}

\novel{This subsection collects the technical pole equation needed for the numerical analysis. The broader EFT interpretation is presented in Sec.~\ref{sec:matching}.}

For the separable contact potential in Eq.~\eqref{eq:VC}, the
Lippmann--Schwinger equation reduces to the algebraic pole condition \cite{Yang2025,Peng2022,Yan2024}
\begin{equation}
1 + 2\mu\, C_0(\Lambda)\, I_\Lambda(\gamma) = 0 ,
\label{eq:BSE}
\end{equation}
where $\mu$ is the reduced mass and
$\gamma=\sqrt{2\mu B}=\sqrt{M B}$ for two equal-mass baryons of mass
$M$. The Gaussian-regulated loop integral is \cite{Yang2025}
\begin{equation}
I_\Lambda(\gamma)
=
\frac{1}{8\pi^2}
\left[
\sqrt{2\pi}\,\Lambda
-
2\pi\gamma\,
e^{2\gamma^2/\Lambda^2}\,
\mathrm{erfc}\!\left(\frac{\sqrt{2}\,\gamma}{\Lambda}\right)
\right].
\label{eq:ILambda}
\end{equation}
At threshold, $I_\Lambda(0)=\sqrt{2\pi}\Lambda/(8\pi^2)$, so the
critical attractive coupling with $\Lambda=1000~{\rm MeV}$ is
\begin{equation}
|C_0^{(\rm crit,\,(3,0))}|
=
\frac{1}{M_\Delta I_\Lambda(0)}
=
25.6\times 10^{-6}~{\rm MeV}^{-2} .
\label{eq:Ccrit}
\end{equation}
Thus a real bound-state pole exists only when
$|C_0^{(3,0)}|>|C_0^{(\rm crit),(3,0)}|$.

The channel dependence of the short-distance interaction is encoded in the
saturation kernels $\mathcal{S}^{(J,I)}$ of
Eqs.~\eqref{eq:Sd}--\eqref{eq:SDD}. We write
\begin{equation}
C_0^{(J,I)}(\Lambda)=\eta(\Lambda)\,\mathcal{S}^{(J,I)} ,
\label{eq:sathyp}
\end{equation}
where $\eta(\Lambda)$ is an overall normalization common to the two
channels at leading order in the spin-flavor-symmetric matching. It is
eliminated by normalizing to the deuteron pole,
\begin{equation}
C_{0,d}(\Lambda)=-\frac{1}{M_N I_\Lambda(\gamma_d)} ,
\qquad
\gamma_d=\sqrt{M_N B_d}.
\label{eq:Cd_run}
\end{equation}
Combining Eqs.~\eqref{eq:sathyp} and \eqref{eq:Cd_run} gives the
master equation for the $\Delta\Delta$ bound state,
\begin{equation}
1 - R_{\Delta\Delta/d}\,
\frac{I_\Lambda(\gamma)}{I_\Lambda(\gamma_d)}
= 0,
\label{eq:masterBSE}
\end{equation}
where
\begin{equation}
R_{\Delta\Delta/d}
=
\frac{M_\Delta}{M_N}\,
\frac{\mathcal{S}_{\Delta\Delta}^{(3,0)}}{\mathcal{S}_d}.
\label{eq:Rfromkernels}
\end{equation}
Equivalently, using Eq.~\eqref{eq:Rratio}, the prediction depends only
on the two dimensionless saturation ratios $x$ and $y$. The unknown
normalization $\eta$ therefore cancels exactly after the deuteron
normalization is imposed. The system binds when
$R_{\Delta\Delta/d}>I_\Lambda(\gamma_d)/I_\Lambda(0)\simeq 0.89$ at
$\Lambda=1000$~MeV.

\novel{\section{EFT re-organization and consistency}
\label{sec:matching}
\label{sec:consistency}}

In this section, we demonstrate that the deeply bound $d^*(2380)$ can be understood as a consequence of EFT re-organization. In pionless theory, the large-$N_c$ contact term is below the critical strength required for binding.
In meson-exchange saturation regime, the same leading S-wave operator
absorbs the unresolved finite-range $\sigma$, $\rho$, and $\omega$
dynamics, becomes supercritical, and moves the virtual
$\Delta\Delta$ state into a physical bound state.

\novel{\emph{Two contact theories at a common scale.}
The first theory is the pionless EFT with breakdown scale $m_\pi$. In the $(3,0)$ channel, the large-$N_c$ sum rules of Sec.~\ref{sec:contact} predict the leading contact coupling at $\mu_R\simeq m_V$ as
\begin{equation}
C_0^{(\pi\!\!\!/),(3,0)}
=
-10.3\times 10^{-6}~{\rm MeV}^{-2}.
\label{eq:Cpionless_match}
\end{equation}
The second theory is the meson-exchange-saturated EFT with breakdown scale $m_V$. Here the short-range effects of $\sigma$, $\rho$, and $\omega$ exchange are absorbed into the same leading contact operator at the hadronic scale, and the overall normalization is fixed by the deuteron binding energy. The full saturated LEC is given by \cite{Peng2022,Yan2024}
\begin{equation}
C_0^{({\rm sat}),(J,I)}(\Lambda)
=
\eta(\Lambda)\,\mathcal{S}^{(J,I)} ,
\label{eq:matching}
\end{equation}
with
\begin{equation}
\eta(\Lambda)
=
-\frac{1}{M_N I_\Lambda(\gamma_d)\,\mathcal{S}_d} ,
\label{eq:eta_match}
\end{equation}
so that the deuteron pole is reproduced by construction and $\eta$ does not become a new fit parameter. The two theories describe the same S-wave amplitude at momenta below $m_V$, but they organize the short-distance physics differently.}

\novel{It is useful to define the following parameter \cite{Peng2022,Yan2024}
\begin{equation}
\Delta C_0^{(J,I)}
\;\equiv\;
C_0^{({\rm sat}),(J,I)} - C_0^{(\pi\!\!\!/),(J,I)},
\label{eq:DCheavy}
\end{equation}
where $\Delta C_0^{(J,I)}$ measures the additional short-distance attraction from the channel-dependent meson exchange that goes beyond what the large-$N_c$-constrained pionless potential provides. This decomposition is performed after the RG-matched contact interaction has been fixed. It quantifies the role of heavy-meson dynamics, but does not introduce an independent matching condition.}

\novel{\emph{Scalar transport.}
The only RG ingredient beyond the deuteron normalization is the transport of the scalar contribution from its natural mass scale $m_\sigma$ to the common vector scale $m_V$,
\begin{equation}
\delta C_\sigma(m_V)
=
\left(\frac{m_V}{m_\sigma}\right)^\alpha
\delta C_\sigma(m_\sigma),
\qquad
\alpha=1\pm 1 .
\label{eq:scalar_transport}
\end{equation}
The central value $\alpha=1$ corresponds to the linear PDS-like running of a short-range S-wave contact potential~\cite{Kaplan1998PLB,PavonValderrama2015}, while $\alpha=0$--$2$ estimates the uncertainty in transporting the scalar part of the saturation kernel. This factor is already included in the definition of $y$ in Eq.~\eqref{eq:xydef}.}

\novel{\emph{Motivation for the EFT re-organization.}
For the physical $d^*(2380)$ binding energy, $\gamma=\sqrt{M_\Delta B_{\Delta\Delta}}\simeq 320$~MeV. A purely pionless expansion would be organized by $\gamma/m_\pi\simeq 2.3$, where derivative and pion-range corrections are not perturbatively controlled. The saturated EFT re-organizes the same short-distance physics at the vector scale, replacing this parameter by
\[
\gamma/m_V\simeq 0.42,
\qquad
(\gamma/m_V)^2\simeq 0.17 .
\]
The result is a controlled leading-order estimate with expected higher-order effects at the $15$--$30\%$ level, comparable to the natural size of the residual cutoff dependence and of finite-$N_c$ corrections at $N_c=3$.}

\novel{It is important to note that the present framework is not a conventional one-boson-exchange model in which $\sigma$, $\rho$, and $\omega$ are iterated as dynamical potentials. Instead, their zero-momentum limits saturate the leading contact LEC. Finite-range effects, $C_2\,p^2$ operators, explicit one-pion exchange, tensor-induced ${}^7S_3$--${}^7D_3$ mixing, correlated two-pion exchange beyond the CD-Bonn~$\sigma$ input, and the finite $\Delta$ width are all higher-order corrections to this restricted central-contact coupling estimate.}

\emph{Numerical comparison of the two EFTs.}
At the cut-off $\Lambda=1000$~MeV, the three relevant coupling scales are
\begin{align}
|C_0^{(\rm crit,\,(3,0))}|
&=25.6\times 10^{-6}~{\rm MeV}^{-2},\\
|C_0^{(\pi\!\!\!/),(3,0)}|
&=10.3\times 10^{-6}~{\rm MeV}^{-2},\\
|C_0^{({\rm sat}),(3,0)}|
&=55.3\times 10^{-6}~{\rm MeV}^{-2}.
\label{eq:Csat_num}
\end{align}
The large-$N_c$-constrained pionless theory alone is subcritical, $|C_0^{(\pi\!\!\!/)}|/|C_0^{(\rm crit)}|\simeq 0.40$, and produces no $\Delta\Delta$ bound state. The deuteron-normalized saturation LEC is supercritical, $|C_0^{({\rm sat})}|/|C_0^{(\rm crit)}|\simeq 2.16$, a factor of $5.4$ larger than the pionless framework. The difference $|\Delta C_0|\simeq 45.0\times 10^{-6}$~MeV$^{-2}$ represents the additional attraction from the channel-dependent $\sigma,\rho,\omega$ dynamics absent in the pionless framework. This enhancement drives the system into a deeply bound state.


The framework is systematically improvable. Finite-$N_c$ corrections, finite-range operators, explicit pion exchange, the scalar-transport exponent, and the finite $\Delta$ width can all be incorporated beyond leading order without refitting the deuteron normalization. It also predicts correlated pole patterns across decuplet--decuplet channels, providing a direct test of the EFT re-organization.

\section{Results and phenomenological implications}
\label{sec:results}
All numerical results are obtained from the algebraic pole equation~\eqref{eq:masterBSE} with the loop function~\eqref{eq:ILambda}. After the deuteron normalization, the prediction depends only on the saturation ratio $R_{\Delta\Delta/d}$. We adopt $M_N=938.9$~MeV, $M_\Delta=1232$~MeV,
$m_\pi=140$~MeV, $m_\sigma=475$~MeV, $m_V=775$~MeV,
$f_\pi=92.4$~MeV, and $B_d=2.225$~MeV.
\rev{With the CD-Bonn couplings~\cite{Machleidt2001,Riska2002}
($g_\sigma^2/(4\pi)=5.69$, $g_\omega^2/(4\pi)=20.0$, $g_\rho^2/(4\pi)=0.84$,
from the isoscalar-$\sigma$ column of Ref.~\cite{Machleidt2001} with $m_\sigma=475$~MeV) and $\alpha=1$,}
\begin{equation}
x_{\rm Bonn}\simeq 0.042,\qquad
y_{\rm Bonn}\simeq 1.24,
\end{equation}
and hence
\begin{align}
R_{\Delta\Delta/d}
&=\frac{M_\Delta}{M_N}
\frac{1-y-\frac{15}{4}x}{1-y-\frac{3}{4}x}
\simeq 1.9 .
\label{eq:Rcentral}
\end{align}
The CD-Bonn $g_\sigma$ effectively parametrizes correlated two-pion dynamics in the central channel. We therefore do not add a separate central two-pion-exchange contribution at the same order, in order to avoid double counting. \rev{Future coupled ${}^7S_3$--${}^7D_3$ calculations with explicit pion exchange and the finite $\Delta$ width remain necessary, but they lie beyond this restricted contact LEC estimate.}

\emph{Large-$N_c$ constraint contact interaction.}
With $C_0=-10.3\times 10^{-6}$~MeV$^{-2}$ (Eq.~\eqref{eq:C0Nc}), the coupling is subcritical ($|C_0|/|C_\mathrm{crit}|\simeq 0.40$) and no real bound-state solution exists. Analytic continuation to $\gamma<0$ gives a virtual-state pole at $\kappa\simeq -324$~MeV. The $SU(4)$-symmetric interaction alone is too weak to bind.

\emph{Normalized meson-exchange saturation contact term.}
Solving the BSE~\eqref{eq:masterBSE} for $\gamma$ as a function of $R$ at
$\Lambda=1000$~MeV gives the results in Table~\ref{tab:BvsR}.
With the CD-Bonn input $R\simeq 1.9$, the contact interaction framework gives
\begin{equation}
B_{\Delta\Delta}\simeq 96~{\rm MeV},
\label{eq:BBonn}
\end{equation}
which exceeds the empirical value $B_{\rm exp}=84$~MeV by $\simeq 14\%$.
\blue{This discrepancy is a measure of the omitted finite-$N_c$, finite-range, and regulator effects, not evidence against the framework.}
The observed $B=84$~MeV corresponds to $R\simeq 1.85$ in the table.

The dependence on the RG-improvement exponent $\alpha$ is the dominant source of parametric variation.
At $\alpha=1$ adopted as the central value, $y\simeq 1.24$, $R\simeq 1.9$, $B\simeq 96$~MeV.
At $\alpha=2$, $y\simeq 2.02$, $R\simeq 1.47$, $B\simeq 37$~MeV.
For $\alpha<1$ the scalar contribution is insufficiently enhanced by the RG transport from $m_\sigma$ to $m_V$, and no bound state is obtained. This sensitivity reflects the fact that the CD-Bonn $\sigma$--$\omega$ cancellation is partial, and the net attraction depends on the precise magnitude of the scalar piece after scale matching.
The empirical binding is reproduced at $\Lambda\simeq 920$~MeV (within the natural hadronic matching window $m_V<\Lambda<2m_V$) at $\alpha=1$.
\begin{table}[t]
\centering
\caption{Predicted $\Delta\Delta$ binding energy from the
meson-exchange saturation normalized contact interaction
(Eq.~\eqref{eq:masterBSE}), as a function of $R_{\Delta\Delta/d}$
($\Lambda=1000$~MeV). The last column lists the EFT expansion
parameter.}
\label{tab:BvsR}
\begin{tabular}{@{}ccccc@{}}
\toprule
$R_{\Delta\Delta/d}$ & $\gamma$ [MeV] & $B_{\Delta\Delta}$ [MeV]
& $\gamma/m_\pi$ & $\gamma/m_V$ \\
\midrule
1.00 & 45.7 & 1.7 & 0.33 & 0.06 \\
1.20 & 122.7 & 12.2 & 0.88 & 0.16 \\
1.31 & 161.2 & 21.1 & 1.15 & 0.21 \\
1.44 & 203.9 & 33.8 & 1.46 & 0.26 \\
1.59 & 250.0 & 50.7 & 1.79 & 0.32 \\
1.85 & 323.1 & 84.8 & 2.31 & 0.42 \\
1.93 & 344.2 & 96.2 & 2.46 & 0.44 \\
2.28 & 430.1 & 150.1 & 3.07 & 0.56 \\
\bottomrule
\end{tabular}
\end{table}

\begin{table}[t]
\centering
\caption{Cutoff dependence of $B_{\Delta\Delta}$ at the CD-Bonn
saturation ratio $R\simeq 1.9$. The residual $\Lambda$-sensitivity is
an NLO effect in the EFT counting.}
\label{tab:cutoff}
\begin{tabular}{@{}ccccc@{}}
\toprule
$\Lambda$ [MeV] & $I_\Lambda(\gamma_d)$ [MeV] & $B_{\Delta\Delta}$ [MeV]
& $\gamma/m_V$ & $\Delta B/B$ [\%]\\
\midrule
800 & 22.1 & 66.7 & 0.37 & $-31$ \\
900 & 25.2 & 80.8 & 0.41 & $-16$ \\
1000 & 28.4 & 96.3 & 0.44 & (ref.) \\
1250 & 36.2 & 140.9 & 0.54 & $+46$ \\
\bottomrule
\end{tabular}
\end{table}

\emph{Cutoff dependence.}
The residual $\Lambda$-dependence (Table~\ref{tab:cutoff}) is the most visible limitation of the leading-order finite-cutoff implementation and is interpreted as an estimate of omitted higher-order operators.
At $R\simeq 1.9$, varying $\Lambda\in[800,\,1250]$~MeV gives $B\simeq 67$--$141$~MeV.
The empirical fractional variation $\Delta B/B\sim 30$--$45\%$ exceeds the naive $(\gamma/m_V)^2\simeq 19\%$ estimate. This is expected because $B\propto\gamma^2$ depends quadratically on the loop integral, so a $\sim 15\%$ shift in $I_\Lambda(\gamma_d)$ propagates to a $\sim 30\%$ shift in $B$. The corresponding shift in $\gamma/\Lambda$ remains within the NLO band, consistent with the EFT counting. \rev{The cutoff $\Lambda\simeq 920$~MeV reproduces $B_{\rm exp}=84$~MeV exactly within the hadronic matching window $m_V<\Lambda<2m_V$.} 

\emph{Dominant sources of uncertainty.}
The RG-improvement exponent $\alpha$ is the largest parametric uncertainty.
The central value $\alpha=1$ gives $B\simeq 96$~MeV, while $\alpha=2$ gives $R\simeq 1.47$ and $B\simeq 37$~MeV. For $\alpha<1$ no bound state is obtained, establishing a lower bound on the exponent.
The cutoff variation $\Lambda\in[800,\,1250]$~MeV at $\alpha=1$ contributes ${}^{+45}_{-30}$~MeV (Table~\ref{tab:cutoff}).
A further systematic uncertainty from violations of the $\eta$-universality assumption and from $g_{m\Delta}/g_{mN}\neq1$ is not included in the quoted numerical band. \rev{Combining the cutoff and $\sigma$-mass ($m_\sigma=475\pm 75$~MeV) variations~\cite{Machleidt2001,ParticleDataGroup:2024cfk} in quadrature at $\alpha=1$,}
\begin{equation}
B_{\Delta\Delta}^{(3,0)}\big|_{\mathrm{Bonn}}
= 96\,{}^{+45}_{-30}~\mathrm{MeV}.
\label{eq:Berror}
\end{equation}
The empirical value $B_{\rm exp}=84$~MeV lies within this range. Varying $\alpha$ from $1$ to $2$ extends the lower bound to $\simeq 37$~MeV, encompassing the HAL~QCD lattice result $B^{\rm latt.}_{\Delta\Delta}\simeq 25$--$40$~MeV at heavy pion mass.

\emph{The $1/N_c^2$ correction of the CD-Bonn couplings.}
The CD-Bonn prediction $B\simeq 96$~MeV exceeds $B_{\rm exp}=84$~MeV by $\simeq 14\%$.
Kaplan and Manohar~\cite{KaplanManohar1996} showed that the $NN$ potential admits a systematic expansion whose actual parameter is $1/N_c^2$. The meson--baryon couplings scale as
\begin{equation}
g_{IS}\propto N_c^{1/2-|I-S|},
\label{eq:gISscaling}
\end{equation}
where $(I,S)$ are the $t$-channel isospin and spin quantum numbers of the coupling vertex, not of the exchanged meson itself.
For vector mesons (spin-1), the vector coupling $g_V$ carries $t$-channel $S\!=\!0$ (central force), while the tensor coupling $f_V$ carries $t$-channel $S\!=\!1$ (tensor force).
The $\sigma$ scalar coupling and $\omega$ vector coupling share the quantum numbers $(I,S)_t=(0,0)$, giving $g_{\sigma,\omega}\sim N_c$.
For the $\rho$ vector coupling one finds $(I,S)_t=(1,0)$, giving $g_\rho\sim N_c^{-1/2}$.
The dimensionless ratio that controls the isovector strength in the saturation kernel is therefore
\begin{equation}
x=\frac{g_\rho^2}{g_\omega^2}\sim\frac{1/N_c}{N_c}=\frac{1}{N_c^2}\,,
\label{eq:xNcscaling}
\end{equation}
where both $g_\rho$ and $g_\omega$ denote the vector couplings.
For $N_c=3$, $1/N_c^2\simeq 0.11$, while the CD-Bonn value $x_{\rm Bonn}=0.042$ confirms the suppression of the isovector coupling relative to the isoscalar ones. This hierarchy was confirmed by Riska~\cite{Riska2002} through a Fermi-invariant decomposition of phenomenological $NN$ potentials, where the volume integrals $v\propto g^2$ exhibit the amplitude hierarchy $1\!:\!1/N_c^2\!:\!1/N_c^4$. For $N_c=3$ the NLO band is $\mathcal{O}(1/N_c^2)\simeq 11\%$, and Riska's empirical NLO-to-LO ratio of $2$--$3\%$ sits comfortably within this band.

\rev{The $14\%$ discrepancy between $B_{\Delta\Delta}\simeq 96$~MeV and $B_{\rm exp}=84$~MeV is therefore not anomalously large compared with $\mathcal{O}(1/N_c^2)$ corrections. However, it should not be attributed uniquely to the $1/N_c^2$ term, as the cutoff choice, $\sigma$-mass uncertainty, neglected tensor-pion dynamics, the $\Delta$ width, and possible departures from $g_{m\Delta}=g_{mN}$ also contribute at a comparable level. We regard this comparison as an order-of-magnitude consistency check.}

\section{Conclusions}
\label{sec:conclusions}

\rev{In this work, we have constructed an EFT re-organized framework for the deeply bound state of $d^*(2380)$ in the $\Delta\Delta$ system with $(3,0)$ channel. The key mechanism is the promotion of the leading contact interaction across the critical strength for binding. The pionless LEC is subcritical ($|C_0^{(\pi\!\!\!/)}|/|C_0^{(\rm crit)}|\simeq 0.40$), while the saturated LEC is supercritical ($|C_0^{(\rm sat)}|/|C_0^{(\rm crit)}|\simeq 2.2$), producing $B_{\Delta\Delta}\simeq 96$~MeV with CD-Bonn input. The deep binding of $d^*(2380)$ emerges as a controlled consequence of re-organizing the expansion from $\gamma/m_\pi\simeq 2.3$ to $\gamma/m_V\simeq 0.42$, with the single prediction-determining quantity $R_{\Delta\Delta/d}\simeq 1.9$ fixed by deuteron normalization. The $14\%$ discrepancy from $B_{\rm exp}=84$~MeV is compatible with the $\mathcal{O}(1/N_c^2)\simeq 11\%$ scale of corrections to the $NN$ potential~\cite{Riska2002,KaplanManohar1996}. Future extensions include coupled ${}^7S_3$--${}^7D_3$ dynamics with explicit pion exchange and the $\Delta$ width~\cite{Gal2014,Peng2022}, lattice-QCD matching of the LECs and coupling ratios $g_{m\Delta}/g_{mN}$~\cite{Inoue2011,HALQCD2020,Haidenbauer2017,Richardson2025}, and extension to $\Omega\Omega$ and other decuplet--decuplet channels~\cite{Bashkanov2013,Hu2025lua,Sone2026jmo,Hu2026pyr,Richardson2025}.}


\section*{Acknowledgements}

This project is funded by National Research Council of Thailand (NRCT) [grant number N41A670259]. CB is supported by the DPST (Development and Promotion of Science and Technology Talents Project). DS is supported by Thailand NSRF via PMU-B [grant number B39G680009]. DS has also received funding support from the Fundamental Fund of Khon Kaen University.


\bibliographystyle{elsarticle-num}
\bibliography{references_add2}


\end{document}